# The Non-Perturbative Quantum Nature of the Dislocation-Phonon Interaction

Mingda Li[1*], Zhiwei Ding[1], Qingping Meng[2], Jiawei Zhou[1], Yimei Zhu[2], Hong Liu[3], M. S. Dresselhaus[3, 4] and Gang Chen[1†]

[1]*Department of Mechanical Engineering, MIT, Cambridge, MA 02139, USA*
[2]*Condensed Matter Physics and Material Sciences Department, Brookhaven National Lab. Upton NY 11973, USA*
[3]*Department of Physics, MIT, Cambridge, MA 02139, USA*
[4] *Department of Electrical Engineering and Computer Sciences, MIT, Cambridge, MA 02139, USA*

Despite the long history of dislocation-phonon interaction studies, there are many problems that have not been fully resolved during this development. These include an incompatibility between a perturbative approach and the long-range nature of a dislocation, the relation between static and dynamic scattering, and the nature of dislocation-phonon resonance. Here by introducing a fully quantized dislocation field, the "dislon"[1], a phonon is renormalized as a quasi-phonon, with shifted quasi-phonon energy, and accompanied by a finite quasi-phonon lifetime that is reducible to classical results. A series of outstanding legacy issues including those above can be directly explained within this unified phonon renormalization approach. In particular, a renormalized phonon naturally resolves the decades-long debate between dynamic and static dislocation-phonon scattering approaches.

It is well known that phonons are strongly scattered by crystal dislocations, resulting in dislocation-induced thermal resistivity [2]. However, after over a half-century long dislocation-phonon interaction (DPI) research, starting from the pioneer studies of static anharmonic scattering by Klemens [3] and Carruthers [4], the understanding of DPI is still not entirely satisfactory [5], leaving behind a series of mysteries to be clarified.

The first problem deals with the degree of validity of the perturbation theory, in particular the Born approximation. This issue was pointed out by Carruthers himself as *possibly invalid but still desirable before more sophisticated calculation* [4], but was often neglected in later developments. To remedy the large quantitative disagreement between Carruthers' theory and thermal conductivity measurements, such as in a prototype material LiF [6], later developments gradually adopted an alternative dynamic scattering mechanism [7-9]. However, there is another possibility that the weak DPI comes from the perturbative analysis procedure other than from static strain scattering itself [4]- A weakly interacting approximation such as the Born approximation may underestimate the dislocation induced thermal resistivity. In fact, the Born approximation breaks down in treating the DPI due to the divergence caused by dislocation's long-range strain field (See Supplemental Material A). To the best of our knowledge, a non-perturbative approach has not been implemented in DPI studies to truly capture the long-range nature of this interaction.

The unsatisfactory early developments triggered a second problem, which is a decades-long debate regarding the origin of dislocation-phonon interaction, namely static strain field scattering [3,10-13] or dynamic vibrating dislocation scattering [14-19]. This debate is partly due to the similar temperature dependence of the thermal conductivity $k \propto T^{2\sim3}$ in either scenario, and such dependence is further limited by the experimental choice of measuring temperature dependence $k(T)$. When dynamic scattering occurs, a dislocation starts to vibrate by absorbing a phonon, and subsequently emit a phonon (Fig. 1a). A consensus gradually formed that a type of viscous dynamic scattering called fluttering plays a significant role in DPI [5,7,9,19,20]. However, the relationship between the static and dynamic scattering is still unclear. In fact, a formulism being able to treat both from equal footing was still unavailable.

Thirdly, the dislocation-phonon resonant scattering process can well be described by the Granato-Lücke model where a dislocation is treated as a vibrating string [16,21,22]. However, this simplification fails to describe many features of resonance, such as the phonon polarization dependence, anisotropy, and long-range strain effect, etc.[5,23]. In particular, because of the restrictive framework of a classical string, most of the studies until now are limited to the classical elastic wave-dislocation scattering mechanism without referring to a quantized phonon [15,23-25].

In this study, we show that a quantum field theory of DPI, based on a quantized field of dislocations, called a "dislon" [1], can easily resolve all above-mentioned problems beyond all expectations (and solve a few other problems as described below). To avoid the uncontrolled expansion in perturbative analysis, an exact functional integral approach is applied, resulting in a renormalized quasi-phonon. The quasi-phonon naturally unifies static and dynamic DPI as the same origin of the dislon field, whose imaginary part can be reduced to the well-known DPI relaxation time.

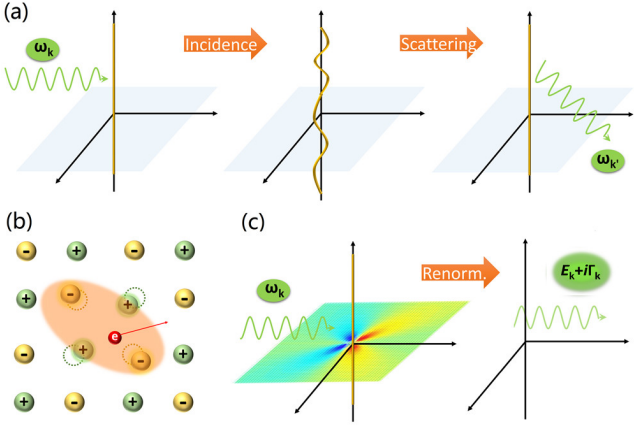

**Fig. 1.** (a) The schematics of classical dynamic dislocation-phonon scattering, where scattering is accomplished when the dislocation absorbs an incoming phonon $\omega_\mathbf{k}$ and re-emits another phonon $\omega_{\mathbf{k}'}$. (b) Due to the electron-ion drag force, an electron is renormalized to a quasi-particle called a "polaron". (c) The quantum picture of the dislocation-phonon interaction. Due to the long-range field of the dislocation, a phonon $\omega_\mathbf{k}$ starts to interact with the dislocation even far away from the core region, making renormalization a more suitable picture than scattering. After renormalization, the strong DPI disappears. We are left with the weakly-interacting quasi-phonons with a renormalized energy $E_\mathbf{k}$ and a finite lifetime $\Gamma_\mathbf{k}$.

In fact, before formal introduction, classical DPI studies have already hinted at phonon renormalization, due to the drag force nature in the DPI [17,22,26-30]. This can be better understood from a direct comparison with a polaron [31]. An electron moving in materials coupling with a phonon will induce a local polarization known as the polaron (Fig. 1b). In the case of the Fröhlich large polaron, the electron-phonon coupling matrix $M(\mathbf{q})$ has component $\propto v_\mathbf{q}$ where $v_\mathbf{q}$ is the electron group velocity [31]. In the case of DPI, due to the similar dragging interaction proportional to the velocity (Eq. 2), it is not difficult to foresee that a phonon can also be renormalized as a quasi-phonon (Fig. 1c). Unlike the classical picture in which a phonon collides with a dislocation and then is being scattered (Fig. 1a), this phonon experiences the long-rage dislocation strain field even far away from the dislocation core. This gives the heuristic reason that a quantum field theory of dislocation capable of describing both its vibrating feature and long range spatial distribution is more suitable to describe the DPI process than classical particle-like scattering due to the extended nature of the dislocation.

To study the quantum interaction between a phonon and a dislocation, we adopt a fully quantized field of a dislocation, "dislon" defined in [1]. A dislon is a quantized collective excitation of a dislocation with vibration and strain energy, where the dislocation's definition $\oint d\mathbf{u} = -\mathbf{b}$ is maintained (See Supplemental Material B). Starting from the second quantized Hamiltonian of the phonon and dislon fields, the total Hamiltonian can be written as

$$H = H_{ph} + H_D + H_{int}$$
$$= \sum_{\mathbf{k}\lambda} \omega_{\mathbf{k}\lambda}\left(b^+_{\mathbf{k}\lambda}b_{\mathbf{k}\lambda} + \frac{1}{2}\right) + \sum_{\kappa} \Omega(\kappa)\left[a^+_\kappa a_\kappa + \frac{1}{2}\right] + H_{int} \quad (1)$$

where $a_\kappa$ and $b_{\mathbf{k}\lambda}$ are dislon and phonon field operators with dispersion $\Omega(\kappa)$ and $\omega_{\mathbf{k}\lambda}$, respectively, and $\kappa \equiv k_z$ since the dislocation line is chosen along the z-direction.

The interaction between the dislocation and phonon originates from the fact that the total lattice displacement $\mathbf{u}_{tot}$ is the vector addition of phonon displacement $\mathbf{u}_{ph}$ and dislocation displacement $\mathbf{u}_{dis}$, i.e. $\mathbf{u}_{tot} = \mathbf{u}_{ph} + \mathbf{u}_{dis}$. This gives kinetic energy cross term $\dot{\mathbf{u}}_{ph}\dot{\mathbf{u}}_{dis}$, harmonic potential cross term $\mathbf{u}_{ph}\mathbf{u}_{dis}$ and anharmonic terms $\mathbf{u}_{ph}^2\mathbf{u}_{dis}^1$ and $\mathbf{u}_{ph}^1\mathbf{u}_{dis}^2$. Since the anharmonic terms only dominate at high temperature and the harmonic potential cross term vanishes [32,33], the dominant term at low temperature is widely accepted to be $\dot{\mathbf{u}}_{ph}\dot{\mathbf{u}}_{dis}$ [5], called fluttering [20]. The corresponding interacting DPI Hamiltonian for a single-mode phonon can be written as (Supplemental Material D)

$$H_{int} = \rho\int \dot{\mathbf{u}}_{ph}(\mathbf{R})\cdot\dot{\mathbf{u}}_D(\mathbf{R})d^3\mathbf{R}$$
$$= \sum_\mathbf{k} \frac{\hbar}{2L}\sqrt{\frac{\rho\omega_{\mathbf{k}\lambda}\Omega(\kappa)}{m(\kappa)}}\left(\boldsymbol{\varepsilon}^*_{\mathbf{k}\lambda}\cdot\mathbf{F}(\mathbf{k})\right)\left(-b^+_\mathbf{k} + b_{-\mathbf{k}}\right)\left(a^+_{-\kappa} - a_\kappa\right) \quad (2)$$

where $\rho$ is the mass density, $\boldsymbol{\varepsilon}^*_{\mathbf{k}\lambda}$ is the phonon polarization vector, $\mathbf{F}(\mathbf{k})$ and $m(\kappa)$ are parameters defined in Supplemental Material D. By rewriting a dislocation line as an extended quantized field, the static dislocation feature is already incorporated since the static case becomes a special case of a full quantized dynamic field. Or rigorously, the definition $\oint d\mathbf{u} = -\mathbf{b}$ is maintained through the dislon field definition, treating the static strain and vibration on an equal footing [1].

Eqs. (1) and (2) can be solved by using an infinite-order Green's function method (See Supplemental Material G), but is too complicated to see essential physics. To gain more physical intuition on the influence of dislocations on phonons but avoid the uncontrolled uncertainty arising from low-order perturbative analysis, we take a non-perturbative functional integral approach [34]. The actions of the non-interacting phonon and dislon field in Eq. (1) in the Matsubara frequency domain can be written as (See Supplemental Material C)

$$S_{ph}(\bar{\phi},\phi) = \sum_{n\mathbf{k}} \bar{\phi}_{\mathbf{k}n}\left(-i\omega_n + \omega_\mathbf{k}\right)\phi_{\mathbf{k}n} \equiv \sum_{n\mathbf{k}} \bar{\phi}_{\mathbf{k}n}G^{-1}_{0n\mathbf{k}}\phi_{\mathbf{k}n}$$
$$S_D(\bar{\chi},\chi) = \sum_{n\kappa} \bar{\chi}_{\kappa n}\left(-i\omega_n + \Omega(\kappa)\right)\chi_{\kappa n} \quad (3)$$

where $\omega_n \equiv 2\pi n k_B T$ is the Matsubara frequency, $\phi_{\mathbf{k}n}$ and

$\chi_{\kappa n}$ are the phonon and dislon field, respectively. The DPI action can be written as (See Supplemental Material D)

$$S_{\text{int}} = \sum_{n\mathbf{k}} \frac{1}{2L} \sqrt{\frac{\rho \omega_\mathbf{k} \Omega(\kappa)}{m(\kappa)}} \boldsymbol{\varepsilon}_\mathbf{k}^* \cdot \mathbf{F}(\mathbf{k}) \begin{pmatrix} -\overline{\phi}_{\mathbf{k}n} \left( \chi_{\kappa n} - \overline{\chi}_{-\kappa,-n} \right) \\ +\phi_{-\mathbf{k}n} \left( \overline{\chi}_{-\kappa n} - \chi_{\kappa,-n} \right) \end{pmatrix} \quad (4)$$

In order to eliminate the dislon degree of freedom, the effective phonon action is defined by integrating out the dislon degree of freedom as

$$S_{\text{eff}}(\overline{\phi}, \phi) \equiv S_{\text{ph}}(\overline{\phi}, \phi) - \log\left[ \int D(\overline{\chi}, \chi) e^{-S_D - S_{\text{int}}} \right] \quad (5)$$

The effective action can finally be simplified using Keldysh rotation [35] and matrix operation as (See Supplemental Material E)

$$S_{\text{eff}}(\overline{\varphi}, \varphi) = \sum_{n\mathbf{k}\mathbf{k}'} \frac{1}{2} \overline{\varphi}_{n\mathbf{k}'} \left[ \frac{G_{0n\mathbf{k}}^{-1} \delta_{\mathbf{k}\mathbf{k}'} - 2J_{n\kappa} \delta_{\kappa\kappa'} g_\mathbf{k}^* g_{\mathbf{k}'} +}{\sqrt{G_{0n\mathbf{k}}^{-2} \delta_{\mathbf{k}\mathbf{k}'} + 4\left( J_{n\kappa} \delta_{\kappa\kappa'} g_\mathbf{k}^* g_{\mathbf{k}'} \right)^2}} \right] \varphi_{n\mathbf{k}} \quad (6)$$

with $g_\mathbf{k} = \left( \boldsymbol{\varepsilon}_\mathbf{k}^* \cdot \mathbf{F}(\mathbf{k}) \right) \sqrt{\omega_\mathbf{k}}$, $R(\kappa) = \frac{\rho}{4A} \frac{\Omega(\kappa)}{m(\kappa)}$, $\varphi$ denoting the effective phonon field to distinguish from the bare phonon field $\phi$, where $J_{\kappa n} \equiv \frac{R(\kappa)\Omega(\kappa)}{\omega_n^2 + \Omega^2(\kappa)}$ is the phonon-dislon coupling strength, containing both dynamic scattering $\omega_n \neq 0$ and static scattering $\omega_n = 0$. Equation (6) is the main result of this theory and contains rich physics: a) When $J_{\kappa n} = 0$ (no DPI), it directly reduces to free phonon action Eq. (3). b) The $\delta_{\kappa\kappa'}$ term indicates that only two phonons with identical z-momentum can be coupled by a dislocation, which makes good physical sense. c) It treats static and dynamic scattering from an equal footing. d) Most importantly, Eq. (6) shows that a phonon is dressed to a quasi-phonon $\omega_\mathbf{k} \rightarrow E_\mathbf{k} + i\Gamma_\mathbf{k}$ upon interaction with a dislocation. As a very rough approximation, the diagonal matrix element in Eq. (6) gives the renormalized quasi-phonon energy $E_\mathbf{k}$ while the off-diagonal part gives quasi-phonon relaxation rate $\Gamma_\mathbf{k}$ (See Supplemental Material F),

$$\begin{aligned} E_\mathbf{k} &\sim \omega_\mathbf{k} - 2J_{n\kappa} |g_\mathbf{k}|^2 \\ \Gamma_\mathbf{k} &\sim 2\sum_{\mathbf{k}'} J_{n\kappa} \delta_{\kappa\kappa'} g_\mathbf{k}^* g_{\mathbf{k}'} \begin{cases} \approx (J_0/\pi) b^2 \omega, \text{ static} \\ \propto 1/\omega, \text{ dynamic} \end{cases} \end{aligned} \quad (7)$$

in the acoustic limit. For the static scattering, it is consistent with the classical Carruthers' result [4] $1/\tau_s \approx N_d \gamma^2 b^2 \omega$ where $\gamma$ is the Grüneisen parameter.

For dynamic scattering, for a given dislon excitation at small $\kappa$, it gives $J_{\kappa n} \propto 1/\omega^2$ after analytical continuation. This finally leads to $1/\tau_d \propto 1/\omega$, consistent with the dynamic scattering results where the relaxation rate is independent of the Burgers vector $\mathbf{b}$ [16,19,36].

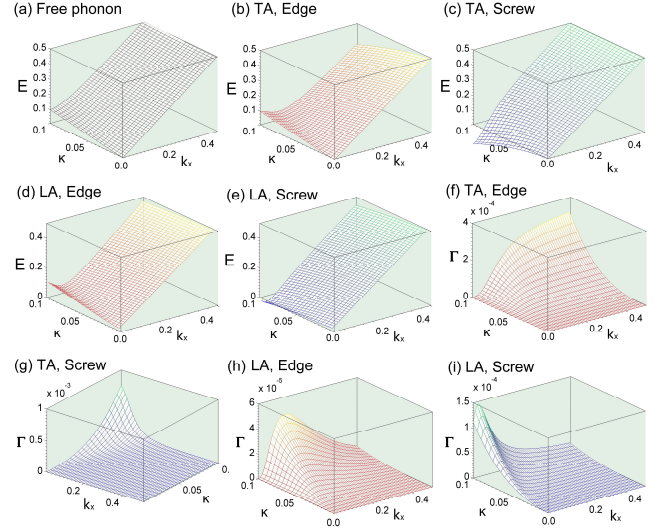

**Fig. 2.** Numerical quasi-phonon spectra when a bare phonon (a) interacts with an edge (red, b, d, f, g) or a screw (blue, c, e, g, i) dislocation. The quasi-phonon energies (b-e) are lowered upon renormalization compared with a bare phonon (a), while the relaxation rates (f-i) show fruitful structures beyond the classical theories.

Eq. (7) also shows a feature of the phonon energy shift. To see this effect in more detail, we diagonalize Eq. (6) numerically to obtain the quasi-phonon spectra (Fig.2 b-e), for an acoustic phonon $\omega_\mathbf{k} = v_s k$ with shear velocity $v_s = 1$, Debye wavenumber $k_D = 1$, sample area $A = 1$ (all are dimensionless for numerical comparison), Poisson ratio $\nu = 0.3$ and xz-plane as slip plane. The renormalized phonon dispersions for either edge dislocation ($\mathbf{b} = [1\ 0\ 0]$, Fig. 2b and d) or screw dislocation ($\mathbf{b} = [0\ 0\ 1]$, Fig. 2c and e), either transverse mode (Fig. 2b and c, $\boldsymbol{\varepsilon}_\mathbf{k} = [0\ 1\ 0]$, $\mathbf{k} \perp \boldsymbol{\varepsilon}_\mathbf{k}$) or longitudinal mode (Fig. 2d and e, $\boldsymbol{\varepsilon}_\mathbf{k} = [1\ 1\ 1]$, $\mathbf{k} \parallel \boldsymbol{\varepsilon}_\mathbf{k}$) demonstrate the feature of red-shift compared with bare phonons (Fig. 2 a).

Intriguingly, compared to the perturbation theories which contain only averaged, structureless simple $\omega$-dependence of relaxation rate $\Gamma_\mathbf{k}$ as in Eq. (7), Eq. (6) has additional power to consider the anisotropic $\mathbf{k}$-dependence hence capture the difference caused by the dislocation type, anisotropy and resonance. Fig. 2 (f-i) shows the relaxation rate $\Gamma_\mathbf{k}$ of edge (Fig. 2 f and h) and screw (Fig. 2 g and i) dislocation with transverse (Fig. 2 f and g) and longitudinal (Fig. 2 h and i) incidence. On one hand, for the transverse instance, at given $\kappa$, $\Gamma_\mathbf{k}$ increases monotonically with $k_x$ for an edge dislocation (Fig. 2f, static-like), but decreases with $k_x$ for a screw dislocation (Fig. 2g, dynamic-like). In other words, even for the same phonon incidence, the DPI can be dominated by dynamic or static scattering depending on dislocation type. On the other hand, for the longitudinal instance, $\Gamma_\mathbf{k}$ shows a peaked value at certain $k$ (Fig. 2 h and i), due to resonant DPI. Such resonance scattering goes far beyond the perturbative approach, where $\Gamma_\mathbf{k}$ can only vary monotonically with $\omega$.

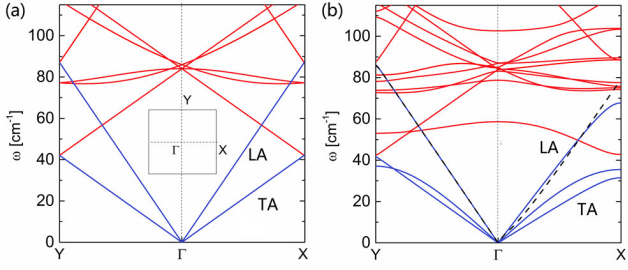

**Fig. 3.** Lattice dynamics simulation of a supercell crystal without (a) and with (b) a dislocation in a supercell Brillouin zone. It is clearly seen that phonon energies are experiencing an anisotropic shift with a drop of group velocity (see for instance, the LA mode). This is highly consistent with the effective theory prediction using Eq. (6).

Phonon spectra are assumed to be unchanged in almost all DPI studies. This is plausible for large samples due to the $1/A$ prefactor in the coupling strength $J_{\kappa n}$, but breaks down for small samples where dislocation has a higher weight. To further validate the prediction of Eq. (6) and to see whether this anisotropic energy renormalization is possibly observable, we performed a lattice dynamics simulation to compute the phonon dispersion with a $30 \times 30 \times 1$ supercell with a hypothetical simple cubic crystal with lattice parameter $a=2.2$ Å, Poisson ratio $\nu = 0$, Young's modulus 540GPa, and creating an edge dislocation with Burgers vector $\mathbf{b} = a\hat{x}$ (Fig. 3). Compared with the non-dislocated dispersion (Fig. 3a), the anisotropic energy red-shift (e.g. no shift along Γ-Y direction, shift from 85 to 70 cm$^{-1}$ for LA mode along Γ-X direction) and reduction of group velocity by lattice dynamics simulation are correctly predicted through effective theory Eq. (6) (black dashed lines in Fig. 3b).

All of the above DPI mechanisms dominate at very low temperature where anharmonic phonon-phonon interaction is weak. Dislocations may also reduce the thermal conductivity above the conductivity maximum temperature [13,16,36,37]. This is explained as contributions beyond the dislocation, such as dislocation dipoles [16] or stacking faults [36]. In this case, the question whether this phenomenon is intrinsic (can be induced by dislocation itself) or extrinsic (such as stacking fault scattering) is not resolved. Here using the quasi-phonon picture, it can be shown directly that a dislocation can actually reduce thermal conductivity at all temperatures. From Eq. (6), The Matsubara Green's function is written as

$$G_{n\mathbf{k}p} \equiv G_{n\mathbf{k},mp} = \langle \varphi_{n\mathbf{k}} \bar{\varphi}_{mp} \rangle = \frac{\delta_{mn} G_{0n\mathbf{k}}}{\frac{1}{2}\delta_{p\mathbf{k}} - J_{n\kappa}G_{0n\mathbf{k}}\delta_{\kappa p_z} g_\mathbf{k}^* g_\mathbf{p} + \sqrt{\frac{1}{4}\delta_{p\mathbf{k}} + \left(J_{n\kappa}G_{0n\mathbf{k}}\delta_{\kappa p_z} g_\mathbf{k}^* g_\mathbf{p}\right)^2}} \quad (8)$$

The corresponding thermal conductivity Kubo formula compatible with Eq. (8) can be computed as (See Supplemental Material H)

$$k(T) = -\frac{k_B \beta^2}{12\pi V} \sum_{\mathbf{kp}} \mathbf{v}_\mathbf{k} \cdot \mathbf{v}_\mathbf{p} E_\mathbf{k} E_\mathbf{p} \int_{-\infty}^{+\infty} d\omega \times \frac{[G_{\mathbf{kp}}^R(\omega) - G_{\mathbf{kp}}^A(\omega)][G_{\mathbf{pk}}^R(\omega) - G_{\mathbf{pk}}^A(\omega)]}{\left[\exp(\beta\omega) - 1\right]^2} \quad (9)$$

where $\mathbf{v}_\mathbf{k} \equiv \partial E_\mathbf{k}/\partial \mathbf{k}$ is the quasi-phonon group velocity, $G^R(\omega)$ and $G^A(\omega)$ are retarded and advanced Green's functions obtained by analytical continuation of Eq. (8). This indicates that the impact of dislocation ranges in all temperatures due to the reduction of the group velocity $\mathbf{v}_\mathbf{k}$. To the best of our knowledge, this is also the first time the change of phonon dispersion caused by dislocation is considered.

The effective theory approach using Eq. (7) resolves another problem related to symmetry breaking. Phonons are well defined within the 1st Brillouin zone as a result of lattice translational symmetry. The long-range strain field caused by dislocation breaks lattice periodicity hence blurs the definition of crystal momentum $\mathbf{k}$ as a good quantum number. In the process of phonon renormalization, according to Eq. (5), the symmetry-breaking dislon field $\chi$ is integrated out, the effect of dislocation on phonon can be discussed using weakly-interacting quasi-phonons within the restored 1st Brillouin zone.

To summarize, we have shown that the dislocation-phonon interaction has the nature of phonon renormalization. This feature is overlooked in perturbative analysis, and only becomes clear through a non-perturbative approach as is adopted here. A renormalized phonon unifies the decades-long debate between static and dynamic dislocation-phonon scattering mechanisms. By treating a dislocation line as a quantized field, both its long range nature and vibrating properties are automatically captured. In the present study, we focus on providing a theoretical framework, but do not intend to seek quantitative agreement with realistic materials, for the following reasons: a) The classical fluttering model [17,18] prior to quantization has already obtained excellent agreement with experiments. b) The relaxation rate is shown to be reducible to classical results. c) Experimental agreement of $k(T)$ is not a sufficient condition to reveal the nature of dislocation induced thermal resistivity, as the **k**-dependence of relaxation is ignored in classical theories, left with a simpler $\omega$-dependence. Instead, we focus on the multiple possibilities of what a quantized dislocation, "dislon", can bring. It opens up an unexplored territory of dislocation-phonon relaxation structure, depending on energy, momentum, anisotropy and dislocation and phonon types. The picture of the renormalized phonon provides insights not only as a conceptual breakthrough, but also as a framework to study the influence of dislocation on material's thermal properties from a fundamental level.

ML would thank W. Cui, J. Mendoza, S. Huang and S. Huberman for their helpful discussions. ML, MSD and GC would like to thank support by S³TEC, an Energy Frontier Research Center funded by U.S. Department of Energy (DOE), Office of Basic Energy Sciences (BES) under Award No. DE-SC0001299/DE-FG02-09ER46577. QM and YZ are supported by DOE-BES, Materials Science and Engineering Division under contract No. DE-SC0012704. HL is supported in part by funds provided by the DOE cooperative research agreement DE-SC0012567.

* mingda@mit.edu
† gchen2@mit.edu

# Supplemental Materials of "The Non-Perturbative Quantum Nature of the Dislocation-Phonon Interaction"

Mingda Li[1], Zhiwei Ding[1], Qingping Meng[2], Jiawei Zhou[1], Yimei Zhu[2], Hong Liu[3], M. S. Dresselhaus[3,4] and Gang Chen[1]

[1]Department of Mechanical Engineering, MIT, Cambridge, MA 02139, USA
[2]Condensed Matter Physics and Material Sciences Department, Brookhaven National Lab. Upton NY 11973, USA
[3]Department of Physics, MIT, Cambridge, MA 02139, USA
[4] Department of Electrical Engineering and Computer Sciences, MIT, Cambridge, MA 02139, USA

## A. The validity of the Born approximation

Heuristically, the Born approximation is valid when the scattered wavefunction $\psi(\mathbf{r})$ does not differ much from the incident wavefunction $\psi_0(\mathbf{r})$. At low-energies, it gives $ma^2|V_0|/\hbar^2 \ll 1$, where $a$ is the potential range and $V_0$ is the potential strength. This is indeed in sharp contradiction with the case of dislocation-phonon interaction, where long-range interaction and strong interaction potential are at present. Quantitatively, the Boltzmann equation describing phonon transport with dislocation line scattering can be written as [S1],

$$\mathbf{v}\cdot\nabla T \frac{dN}{dT} = \frac{1}{V^{2/3}}\left(\frac{gb\alpha}{\rho}\right)^2 \frac{q\left(1-q_z^2/q^2\right)}{24\pi|\mathbf{v}|^3}\int d\phi' F(\phi,\phi')(n_q' - n_q) \quad (A1)$$

Where $\mathbf{v}$ is the average sound velocity, $g$ is a dimensionless constant, $b$ is the Burgers vector, $\alpha \equiv \frac{1-2\nu}{1-\nu}$ with $\nu$ the Poisson ratio, $\rho$ is the mass density. Most importantly, $F(\phi,\phi')$ is a kernel function satisfying $F(\phi,\phi') \propto 1/[1-\cos(\phi-\phi')]$, resulting a divergent pole for forward scattering $\phi \approx \phi'$ hence invalidating the Born approximation. To remedy this divergence, the author has made a bold assumption that $n_q' - n_q = n_q[1-\cos(\phi-\phi')]$ to cancel out the divergence and circumvent the issue from perturbation analysis. This assumption groundlessly links the scattering angular distribution $\phi$ to the Bose occupation factor $n_q$, as mentioned by the author in [S1] as "*hides a multitude of sins*".

## B. Summary of Dislon- the quantized field of crystal dislocation

Assuming that an infinite-long straight dislocation is located at (x,y)=(0,0) along z-direction, with the displacement of the dislocation at position $z$ is $Q(z)$, as shown in [S2]. Then we have

$$Q(z) = \sum_\kappa Q_\kappa e^{i\kappa z} \quad (B1)$$

Defining $\mathbf{u}(\mathbf{R})$ as the displacement at spatial point $\mathbf{R}$, i.e. deviation of lattice point away from the equilibrium position. Then the $i^{th}$ component of $\mathbf{u}$ can be expanded as

$$u_i(\mathbf{R}) = u_i(\mathbf{R} \equiv (\mathbf{r}, z) \equiv (x, y, z)) = \sum_\kappa f_i(\mathbf{r}; \kappa) e^{i\kappa z} Q_\kappa$$
$$= \frac{1}{A} \sum_{\mathbf{s},\kappa} F_i(\mathbf{s}; \kappa) e^{+i\mathbf{s}\cdot\mathbf{r}} e^{i\kappa z} Q_\kappa = \frac{1}{A} \sum_{\mathbf{k}\equiv(\mathbf{s},\kappa)} F_i(\mathbf{k}) e^{i\mathbf{k}\cdot\mathbf{R}} Q_\kappa \quad \text{(B2)}$$

For the phonon case, it is a plane wave expansion. If we recall Fourier transform $\int d^2 r e^{i(\mathbf{s}-\mathbf{s}')\cdot\mathbf{r}} = A\delta_{\mathbf{s},\mathbf{s}'}$ and $\sum_\mathbf{s} e^{i\mathbf{s}\cdot(\mathbf{r}-\mathbf{r}')} = A\delta(\mathbf{r}-\mathbf{r}')$. The coefficients have been computed as

$$F_i(\mathbf{s};\kappa) = +\frac{i}{k^2}\left( n_i(\mathbf{b}\cdot\mathbf{k}) + b_i(\mathbf{n}\cdot\mathbf{k}) - \frac{1}{(1-\nu)} \frac{k_i(\mathbf{n}\cdot\mathbf{k})(\mathbf{b}\cdot\mathbf{k})}{k^2} \right) \quad \text{(B3)}$$

In this way, the classical kinetic energy and potential energy of this dislocation can be written as

$$H = T + U = \frac{L}{2}\sum_\kappa m(\kappa)\dot{Q}_\kappa \dot{Q}_\kappa^* + \frac{L}{2}\sum_\kappa \kappa^2 K(\kappa) Q_\kappa Q_\kappa^* \quad \text{(B4)}$$

with $m(\kappa) \equiv \frac{\rho}{A}\sum_{i=1}^{3}\sum_\mathbf{s}|F_i(\mathbf{s};\kappa)|^2$.

In a quantum-mechanical picture, by imposing canonical quantization condition
$$[Q_\kappa, P_{\kappa'}] = i\delta_{\kappa,\kappa'} \quad \text{(B5)}$$

We could define the creation and annihilation operators by
$$\begin{cases} Q_\kappa = Z_\kappa \left[ a_\kappa + a_{-\kappa}^+ \right] \\ P_\kappa = \frac{i}{2Z_\kappa}\left[ a_\kappa^+ - a_{-\kappa} \right] \end{cases} \quad \text{(B6)}$$

With the conjugate momentum $P_\kappa = \frac{\partial \mathcal{L}}{\partial \dot{Q}_\kappa} = Lm(\kappa)\dot{Q}_\kappa^*$ with $Z_\kappa = \sqrt{\frac{1}{2Lm(\kappa)\Omega(\kappa)}}$. Hence Eq. (B4) for a single dislocation line is rewritten as

$$H_D = \sum_\kappa \Omega(\kappa)\left[ a_\kappa^+ a_\kappa + \frac{1}{2} \right] \quad \text{(B7)}$$

with eigenfrequency $\Omega(\kappa) = \kappa\sqrt{\frac{K(\kappa)}{m(\kappa)}}$.

For an edge dislocation, we have [S2]
$$m_E(\kappa) = \frac{\rho b^2}{4\pi}\left[ \log\left(1 + \frac{k_D^2}{\kappa^2}\right) - \frac{k_D^2}{k_D^2 + \kappa^2} + \frac{4\nu - 3}{8(1-\nu)^2}\left( \log\left(1 + \frac{k_D^2}{\kappa^2}\right) - \frac{k_D^2(3k_D^2 + 2\kappa^2)}{2(k_D^2 + \kappa^2)^2} \right) \right]$$

$$K_E(\kappa) = \frac{\mu b^2}{4\pi}\left[ \frac{1-2\nu}{2(1-\nu)}\log\left(\frac{k_D^2}{\kappa^2} + 1\right) + 1 - \frac{1}{4(1-\nu)}\frac{\kappa^2}{k_D^2 + \kappa^2} \right]$$

$$\Omega_E(\kappa) = v_s \kappa \sqrt{\frac{\frac{1-2\nu}{2(1-\nu)}\log\left(\frac{k_D^2}{\kappa^2}+1\right)+1-\frac{1}{4(1-\nu)}\frac{\kappa^2}{k_D^2+\kappa^2}}{\log\left(1+\frac{k_D^2}{\kappa^2}\right)-\frac{k_D^2}{k_D^2+\kappa^2}+\frac{4\nu-3}{8(1-\nu)^2}\left(\log\left(1+\frac{k_D^2}{\kappa^2}\right)-\frac{k_D^2\left(3k_D^2+2\kappa^2\right)}{2\left(k_D^2+\kappa^2\right)^2}\right)}}$$

And for a screw dislocation, we have

$$m_S(\kappa) = \frac{\rho b^2}{4\pi}\left[\frac{k_D^2}{2(\kappa^2+k_D^2)}+\frac{1}{2}\log\left(1+\frac{k_D^2}{\kappa^2}\right)+\frac{4\nu-3}{4(1-\nu)^2}\frac{k_D^4}{\left(k_D^2+\kappa^2\right)^2}\right]$$

$$K_S(\kappa) = \frac{\mu b^2}{4\pi}\left[\frac{1+\nu}{2(1-\nu)}\left(\log\left(1+\frac{k_D^2}{\kappa^2}\right)-1\right)+\frac{1}{1-\nu}\frac{\kappa^2}{k_D^2+\kappa^2}\right]$$

$$\Omega_S(\kappa) = v_s\kappa \sqrt{\frac{\frac{1+\nu}{2(1-\nu)}\left(\log\left(1+\frac{k_D^2}{\kappa^2}\right)-1\right)+\frac{1}{1-\nu}\frac{\kappa^2}{k_D^2+\kappa^2}}{\frac{k_D^2}{2(\kappa^2+k_D^2)}+\frac{1}{2}\log\left(1+\frac{k_D^2}{\kappa^2}\right)+\frac{4\nu-3}{4(1-\nu)^2}\frac{k_D^4}{\left(k_D^2+\kappa^2\right)^2}}}$$

## C. Derivation of Dislon and phonon actions

From Eq. (B7) and Eq. (1) in main text, as non-interacting Bosonic quasiparticles, the non-interacting phonons and dislon actions in imaginary time can be written using the method in [S3] as

$$S_{ph} = \int_0^\beta d\tau \sum_{\mathbf{k}} \bar{\phi}_{\mathbf{k}}(\tau)\left(\partial_\tau + \hbar\omega_{\mathbf{k}}\right)\phi_{\mathbf{k}}(\tau)$$

$$S_D = \int_0^\beta d\tau \sum_{\kappa} \bar{\chi}_\kappa(\tau)\left(\partial_\tau + \hbar\Omega(\kappa)\right)\chi_\kappa(\tau)$$

(C1)

Where $\phi_{\mathbf{k}}(\tau)$ and $\chi_\kappa(\tau)$ are phonon and dislon fields, respectively, and $\omega_{\mathbf{k}\lambda}$ and $\Omega(\kappa)$ are phonon and dislon dispersion. Defining the phonon and dislon fields using Fourier transformed Matsubara form as

$$\phi_{\mathbf{k}}(\tau) = \frac{1}{\sqrt{\beta}}\sum_n \phi_{\mathbf{k}n}e^{+i\omega_n\tau}, \quad \phi_{\mathbf{k}n} = \frac{1}{\sqrt{\beta}}\int_0^\beta \phi_{\mathbf{k}}(\tau)e^{-i\omega_n\tau}d\tau$$

$$\chi_\kappa(\tau) = \frac{1}{\sqrt{\beta}}\sum_n \chi_{\kappa n}e^{+i\omega_n\tau}, \quad \chi_{\kappa n} = \frac{1}{\sqrt{\beta}}\int_0^\beta \chi_\kappa(\tau)e^{-i\omega_n\tau}d\tau$$

(C2)

with $\omega_n = 2\pi n/\beta$ are the Bosonic Matsubara frequency. Then using $\int_0^\beta e^{-i(\omega_n-\omega_m)\tau}d\tau = \beta\omega_{mn}$, the imaginary-time actions can be rewritten in terms of Matsubara frequency domain as

$$S_{ph} = \sum_{n\mathbf{k}} \bar{\phi}_{\mathbf{k}n}\left(-i\omega_n + \omega_\mathbf{k}\right)\phi_{\mathbf{k}n}$$
$$S_D = \sum_{n\kappa} \bar{\chi}_{\kappa n}\left(-i\omega_n + \Omega(\kappa)\right)\chi_{\kappa n}$$
(C3)

## D. Derivation of dislocation-phonon interaction action

We derive the action from Hamiltonian approach. The interaction Hamiltonian comes from the fact that total displacement $\mathbf{u}_{tot} = \mathbf{u}_{ph} + \mathbf{u}_{dis}$. The cross term from potential energy $\mathbf{u}_{ph} \cdot \mathbf{u}_{dis}$ vanishes [S4, S5], left with the non-vanishing cross-term from kinetic energy:

$$H_{int} = \rho\int \dot{\mathbf{u}}_{ph}(\mathbf{R})\cdot\dot{\mathbf{u}}_D(\mathbf{R})d^3\mathbf{R} = \int \mathbf{p}_{ph}(\mathbf{R})\cdot\dot{\mathbf{u}}_D(\mathbf{R})d^3\mathbf{R}$$
$$= \int \frac{1}{\sqrt{V}}\sum_{\mathbf{k}'}\mathbf{p}_{ph,\mathbf{k}'}e^{i\mathbf{k}'\cdot\mathbf{R}}\cdot\frac{1}{A}\sum_{\mathbf{k}}\mathbf{F}(\mathbf{k})e^{i\mathbf{k}\cdot\mathbf{R}}\dot{Q}_\kappa d^3\mathbf{R} = \frac{1}{\sqrt{L}}\sum_{\mathbf{k}}\mathbf{p}^*_{ph,\mathbf{k}}\cdot\mathbf{F}(\mathbf{k})\dot{Q}_\kappa$$
(D1)

Now using the facts from canonical commutation relation $\dot{Q}_\kappa = \frac{P^*_\kappa}{Lm(\kappa)} = +i\sqrt{\frac{\Omega(\kappa)}{2Lm(\kappa)}}\left(a^+_{-\kappa} - a_\kappa\right)$ and $\mathbf{p}_{ph,\mathbf{k}} = \sum_\lambda i\sqrt{\frac{\rho\omega_{\mathbf{k}\lambda}}{2}}\left(-b_{\mathbf{k}\lambda} + b^+_{-\mathbf{k}\lambda}\right)\boldsymbol{\varepsilon}_{\mathbf{k}\lambda}$ with $\boldsymbol{\varepsilon}_{\mathbf{k}\lambda}$ is the phonon polarization vector, the Hamiltonian can be further re-written in 2$^{nd}$ quantization form as

$$H_{int} = \sum_{\mathbf{k}\lambda}\frac{1}{2L}\sqrt{\frac{\rho\omega_{\mathbf{k}\lambda}\Omega(\kappa)}{m(\kappa)}}\left(\boldsymbol{\varepsilon}^*_{\mathbf{k}\lambda}\cdot\mathbf{F}(\mathbf{k})\right)\left(-b^+_{\mathbf{k}\lambda} + b_{-\mathbf{k}\lambda}\right)\left(a^+_{-\kappa} - a_\kappa\right)$$
(D2)

where $m(\kappa)$ is a parameter with linear mass density defined in Section B, $\rho$ is the mass density of the material, and the coefficient

$$\boldsymbol{\varepsilon}_\mathbf{k}\cdot\mathbf{F}(\mathbf{k}) = +\frac{i}{k^2}\left((\mathbf{n}\cdot\boldsymbol{\varepsilon}_\mathbf{k})(\mathbf{b}\cdot\mathbf{k}) + (\mathbf{b}\cdot\boldsymbol{\varepsilon}_\mathbf{k})(\mathbf{n}\cdot\mathbf{k}) - \frac{1}{(1-\nu)}\frac{(\boldsymbol{\varepsilon}_\mathbf{k}\cdot\mathbf{k})(\mathbf{n}\cdot\mathbf{k})(\mathbf{b}\cdot\mathbf{k})}{k^2}\right)$$
(D3)

which satisfies $\left[\boldsymbol{\varepsilon}_\mathbf{k}\cdot\mathbf{F}(\mathbf{k})\right]^* = \boldsymbol{\varepsilon}_{-\mathbf{k}}\cdot\mathbf{F}(-\mathbf{k})$. From now on we neglect the phonon mode label λ, but only studies the interaction between single-mode phonon and dislocations.

$$S_{int} = \int_0^\beta d\tau\sum_\mathbf{k}\frac{1}{2L}\sqrt{\frac{\rho\omega_\mathbf{k}\Omega(\kappa)}{m(\kappa)}}\left(\boldsymbol{\varepsilon}^*_\mathbf{k}\cdot\mathbf{F}(\mathbf{k})\right)\left(-\bar{\phi}_\mathbf{k}(\tau) + \phi_{-\mathbf{k}}(\tau)\right)\left(\bar{\chi}_{-\kappa}(\tau) - \chi_\kappa(\tau)\right)$$
(D4)

# E. Derivation of phonon effective action

The total partition function is written as functional integrals of both phonon and dislon fields, or equivalently effective phonon field as

$$Z = \int D(\bar{\phi}, \phi) D(\bar{\chi}, \chi) e^{-\left(S_{ph}(\bar{\phi}, \phi) + S_D(\bar{\chi}, \chi) + S_{int}(\bar{\phi}, \phi, \bar{\chi}, \chi)\right)} \equiv \int D(\bar{\phi}, \phi) e^{-S_{eff}(\bar{\phi}, \phi)} \quad (E1)$$

Where $D$ denotes the functional integration over all field configuration.

The effective action of phonon can be defined after elimination of dislon degree of freedom as

$$S_{eff}(\bar{\phi}, \phi) = S_{ph}(\bar{\phi}, \phi) - \log\left(\int D(\bar{\chi}, \chi) e^{-\left(S_D(\bar{\phi}, \phi) + S_{int}(\bar{\phi}, \phi, \bar{\chi}, \chi)\right)}\right) \quad (E2)$$

which is rigorous for weak dislocation-phonon interaction, can be computed using the generalized Gaussian integral [S3]

$$\int D(\bar{\chi}, \chi) e^{-\left(S_D(\bar{\phi}, \phi) + S_{int}(\bar{\phi}, \phi, \bar{\chi}, \chi)\right)} = \int D[\bar{\chi}, \chi] \exp\left[-\sum_{n\kappa}\left[\begin{array}{l}\bar{\chi}_{n\kappa}\left(-i\omega_n + \Omega(\kappa)\right)\chi_{n\kappa} + \\ \sum_{s}\frac{1}{2L}\sqrt{\frac{\rho\omega_k\Omega(\kappa)}{m(\kappa)}}\left(\varepsilon_{-k}^* \cdot F(-k)\right)\left(\phi_{k,n} - \bar{\phi}_{-k,-n}\right)\bar{\chi}_{\kappa n} \\ + \sum_{s}\frac{1}{2L}\sqrt{\frac{\rho\omega_k\Omega(\kappa)}{m(\kappa)}}\left(\varepsilon_k^* \cdot F(k)\right)\left(\bar{\phi}_{k,n} - \phi_{-k,-n}\right)\chi_{\kappa n}\end{array}\right]\right]$$

$$= \exp\left[\sum_{n\kappa}\begin{array}{l}\left[\sum_{s}\frac{1}{2L}\sqrt{\frac{\rho\omega_k\Omega(\kappa)}{m(\kappa)}}\left(\varepsilon_{-k}^* \cdot F(-k)\right)\left(\phi_{k,n} - \bar{\phi}_{-k,-n}\right)\right] \times \frac{1}{-i\omega_n + \Omega(\kappa)} \\ \times\left[\sum_{s'}\frac{1}{2L}\sqrt{\frac{\rho\omega_k\Omega(\kappa)}{m(\kappa)}}\left(\varepsilon_k^* \cdot F(k)\right)\left(\bar{\phi}_{k,n} - \phi_{-k,-n}\right)\right]\end{array}\right]$$

where **s** and **s'** denote the in-plane 2D momentum index before and after scattering, respectively. Defining $\mathbf{k} \equiv (\mathbf{s}, \kappa)$ is the label for 3D momentum integral, and the scattering with dislocation along z couples phonons with different in-plane 2D momentum **s** to **s'**, but the same momentum in z-direction, which is quite reasonable. The above equation can be further reduced to

$$S_{eff}(\bar{\phi}, \phi) = \sum_{n\mathbf{kk'}}\left(\bar{\phi}_{\mathbf{k}n}\left(-i\omega_n + \omega_{\mathbf{k}}\right)\phi_{\mathbf{k}n}\delta_{\mathbf{kk'}} - R(\kappa)\delta_{\kappa\kappa'}g_{\mathbf{k}}^*g_{\mathbf{k'}}\frac{\left(\phi_{\mathbf{k},n} - \bar{\phi}_{-\mathbf{k},-n}\right)\left(\bar{\phi}_{\mathbf{k'}n} - \phi_{-\mathbf{k'},-n}\right)}{-i\omega_n + \Omega(\kappa)}\right) \quad (E3)$$

where $g_{\mathbf{k}} = \left(\varepsilon_{\mathbf{k}}^* \cdot F(\mathbf{k})\right)\sqrt{\omega_{\mathbf{k}}}$, $R(\kappa) = \frac{\rho}{4A}\frac{\Omega(\kappa)}{m(\kappa)}$. Now using the fact $\Omega(-\kappa) = +\Omega(\kappa)$, $\left[\varepsilon_{\mathbf{k}} \cdot F(\mathbf{k})\right]^* = \varepsilon_{-\mathbf{k}} \cdot F(-\mathbf{k})$ hence $g_{\mathbf{k}}^* = g_{-\mathbf{k}}$ valid, we have

$$S_{eff}(\bar{\phi}, \phi) = \sum_{n\mathbf{k}}\bar{\phi}_{\mathbf{k}n}\left(-i\omega_n + \omega_{\mathbf{k}}\right)\phi_{\mathbf{k}n} - \sum_{n\mathbf{kk'}}\frac{\Omega(\kappa)R(\kappa)\delta_{\kappa\kappa'}g_{\mathbf{k}}^*g_{\mathbf{k'}}}{\Omega^2(\kappa) + \omega_n^2}\left(\phi_{\mathbf{k}n} - \bar{\phi}_{-\mathbf{k}-n}\right)\left(\bar{\phi}_{\mathbf{k'}n} - \phi_{-\mathbf{k'}-n}\right) \quad (E4)$$

Now performing Keldysh rotation, that

$$\psi_{1\mathbf{k}n} = \frac{1}{\sqrt{2}}\left(\phi_{\mathbf{k}n} + \bar{\phi}_{-\mathbf{k}-n}\right), \quad \bar{\psi}_{1\mathbf{k}n} = \frac{1}{\sqrt{2}}\left(\bar{\phi}_{\mathbf{k}n} + \phi_{-\mathbf{k}-n}\right)$$

$$\psi_{2\mathbf{k}n} = \frac{1}{\sqrt{2}}\left(\phi_{\mathbf{k}n} - \bar{\phi}_{-\mathbf{k}-n}\right), \quad \bar{\psi}_{2\mathbf{k}n} = \frac{1}{\sqrt{2}}\left(\bar{\phi}_{\mathbf{k}n} - \phi_{-\mathbf{k}-n}\right)$$

(E5)

The action in Eq. (E4) can further be simplified as

$$S_{\text{eff}}(\bar{\psi},\psi) = \sum_{n\mathbf{k}} \frac{(-i\omega_n + \omega_\mathbf{k})}{2} (\bar{\psi}_{1\mathbf{k}n} \ \bar{\psi}_{2\mathbf{k}n}) \begin{pmatrix} 1 & 1 \\ 1 & 1 \end{pmatrix} \begin{pmatrix} \psi_{1\mathbf{k}n} \\ \psi_{2\mathbf{k}n} \end{pmatrix} - \sum_{n\mathbf{k}\mathbf{k}'} \frac{2\Omega(\kappa)R(\kappa)\delta_{\kappa\kappa'} g_\mathbf{k}^* g_{\mathbf{k}'}}{\Omega^2(\kappa)+\omega_n^2} \bar{\psi}_{2\mathbf{k}'n}\psi_{2\mathbf{k}n} \quad (E6)$$

Defining $J_{\kappa n} \equiv \dfrac{R(\kappa)\Omega(\kappa)}{\omega_n^2+\Omega^2(\kappa)}$, $G_{0n\mathbf{k}}^{-1} \equiv -i\omega_n + \omega_\mathbf{k}$, we have the matrix form of effective action

$$S_{\text{eff}}(\bar{\psi},\psi) = \sum_{n\mathbf{k}\mathbf{k}'} (\bar{\psi}_{1\mathbf{k}'n} \ \bar{\psi}_{2\mathbf{k}'n}) \left[ \frac{1}{2} G_{0n\mathbf{k}}^{-1} \delta_{\mathbf{k}\mathbf{k}'} \begin{pmatrix} 1 & 1 \\ 1 & 1 \end{pmatrix} - 2 J_{\kappa n}\delta_{\kappa\kappa'} g_\mathbf{k}^* g_{\mathbf{k}'} \begin{pmatrix} 0 & 0 \\ 0 & 1 \end{pmatrix} \right] \begin{pmatrix} \psi_{1\mathbf{k}n} \\ \psi_{2\mathbf{k}n} \end{pmatrix} \quad (E7)$$

Now performing canonical transformation $\begin{pmatrix} \psi_{1n\mathbf{k}} \\ \psi_{2n\mathbf{k}} \end{pmatrix} \to \begin{pmatrix} \varphi_{1n\mathbf{k}} \\ \varphi_{2n\mathbf{k}} \end{pmatrix}$ to diagonalize the $2\times 2$ matrix, we have

$$S_{\text{eff}}(\bar{\varphi},\varphi)$$

$$= \sum_{n\mathbf{k}\mathbf{k}'} (\bar{\varphi}_{1n\mathbf{k}} \ \bar{\varphi}_{2n\mathbf{k}}) \begin{pmatrix} \dfrac{1}{2} G_{0n\mathbf{k}}^{-1} \delta_{\mathbf{k}\mathbf{k}'} - J_{\kappa n}\delta_{\kappa\kappa'} g_\mathbf{k}^* g_{\mathbf{k}'} \\ -\sqrt{\dfrac{1}{4} G_{0n\mathbf{k}}^{-2} \delta_{\mathbf{k}\mathbf{k}'} + \left( J_{\kappa n}\delta_{\kappa\kappa'} g_\mathbf{k}^* g_{\mathbf{k}'}\right)^2} & 0 \\ 0 & \dfrac{1}{2} G_{0n\mathbf{k}}^{-1} \delta_{\mathbf{k}\mathbf{k}'} - J_{\kappa n}\delta_{\kappa\kappa'} g_\mathbf{k}^* g_{\mathbf{k}'} \\ & +\sqrt{\dfrac{1}{4} G_{0n\mathbf{k}}^{-2} \delta_{\mathbf{k}\mathbf{k}'} + \left( J_{\kappa n}\delta_{\kappa\kappa'} g_\mathbf{k}^* g_{\mathbf{k}'}\right)^2} \end{pmatrix} \begin{pmatrix} \varphi_{1n\mathbf{k}} \\ \varphi_{2n\mathbf{k}} \end{pmatrix} \quad (E8)$$

Notice that when $J_{\kappa n} = 0$, it should be reduced to non-interacting phonon solution instead of 0, we only need the (2, 2) component. Defining $\varphi_{2n\mathbf{k}} \equiv \varphi_{n\mathbf{k}}$, we finally obtain

$$S_{\text{eff}}(\bar{\varphi},\varphi) = \sum_{n\mathbf{k}\mathbf{k}'} \frac{1}{2} \bar{\varphi}_{n\mathbf{k}'} \left[ G_{0n\mathbf{k}}^{-1} \delta_{\mathbf{k}\mathbf{k}'} - 2 J_{n\kappa}\delta_{\kappa\kappa'} g_\mathbf{k}^* g_{\mathbf{k}'} + \sqrt{G_{0n\mathbf{k}}^{-2} \delta_{\mathbf{k}\mathbf{k}'} + 4\left( J_{n\kappa}\delta_{\kappa\kappa'} g_\mathbf{k}^* g_{\mathbf{k}'}\right)^2} \right] \varphi_{n\mathbf{k}} \quad (E9)$$

## F. The consistency of relaxation time

Since Eq. (E9) is not diagonalized in k, we could write it back in terms of $\omega$, and assume a linear dependence $\omega = vk$ (acoustic approximation, $v$ is magnitude of sound velocity). Then we have

$$\boldsymbol{\varepsilon}_\mathbf{k} \cdot \mathbf{F}(\mathbf{k}) \approx \frac{b}{k} = \frac{bv}{\omega}, \quad g_\mathbf{k} \approx \frac{bv}{\sqrt{\omega}} \quad (F1)$$

In the static limit, $\omega_n = 0$, for simplicity we assume $J_{\kappa 0} \approx J_0$ a constant as it is a slow-varying function of $\kappa$. For a given $\mathbf{k}$, due to same z-component, the summation over $\mathbf{k}$ is 2-dimensional. The summation gives

$$\sum_{\mathbf{k'}} g_{\mathbf{k'}} \delta_{\kappa\kappa'} \propto \frac{1}{4\pi^2} \int d^2 k' g_{\mathbf{k'}} \delta_{\kappa\kappa'} \simeq \frac{1}{4\pi^2 v^2} \int d^2\omega g(\omega)$$

$$= \frac{1}{2\pi v^2} \int g(\omega) \omega d\omega = \frac{1}{2\pi v^2} \int \frac{bv}{\sqrt{\omega}} \omega d\omega \simeq \frac{1b}{2\pi v} \omega^{3/2} \tag{F2}$$

Then the coupling term gives

$$\sum_{\mathbf{k'}} 2 J_{n\kappa} \delta_{\kappa\kappa'} g_{\mathbf{k}}^* g_{\mathbf{k'}} \approx 2 J_0 \frac{bv}{\sqrt{\omega}} \frac{b}{2\pi v} \omega^{3/2} = \left(\frac{J_0}{\pi}\right) b^2 \omega \Rightarrow$$

$$\frac{1}{\tau_\omega} = \left(\frac{J_0}{\pi}\right) b^2 \omega \tag{F3}$$

Which is fully consistent with the Carruthers' result in [S1], with the proportionality with $b^2$ and $\omega$, independent with sound velocity $v$.

For dynamic scattering, after analytical continuation, we have $J_{\kappa n} \approx \frac{R(\kappa)\Omega(\kappa)}{\omega^2}$ when $\Omega(\kappa)$ is small, giving

$$\sum_{\mathbf{k'}} 2 J_{n\kappa} \delta_{\kappa\kappa'} g_{\mathbf{k}}^* g_{\mathbf{k'}} \approx 2A \frac{R(\kappa)\Omega(\kappa)}{\omega^2} \frac{bv}{\sqrt{\omega}} \frac{b}{2\pi v} \omega^{3/2} = A \frac{R(\kappa)\Omega(\kappa)}{\pi\omega} b^2 =$$

$$= \frac{\rho}{4\pi\omega} \times \frac{\Omega^2(\kappa)}{m(\kappa)} b^2 \propto \frac{1}{\omega} \tag{F4}$$

Where the explicit $b^2$ dependence is cancelled out from the definition of $m(\kappa)$.

## G. Full diagonalization of quasi-phonon energy and lifetime

A full diagonalization procedure is provided in this section.
The single-mode phonon- dislon full Hamiltonian can be written as

$$H = T + U = \sum_{\mathbf{k}} \omega_{\mathbf{k}} \left( b_{\mathbf{k}}^+ b_{\mathbf{k}} + \frac{1}{2} \right) + \sum_\kappa \Omega(\kappa) \left[ a_\kappa^+ a_\kappa + \frac{1}{2} \right] + H_{int}$$

$$H_{int} = \sum_{\mathbf{k}} \frac{1}{2L} \sqrt{\frac{\rho\omega_{\mathbf{k}}\Omega(\kappa)}{m(\kappa)}} \left( \boldsymbol{\varepsilon}_{\mathbf{k}}^* \cdot \mathbf{F}(\mathbf{k}) \right) \left( -b_{\mathbf{k}}^+ + b_{-\mathbf{k}} \right) \left( a_{-\kappa}^+ - a_\kappa \right) \tag{G1}$$

Now performing canonical transform, that defining operators

$$A_k = \frac{1}{\sqrt{2}} \left( b_k + b_{-k}^+ \right) = A_{-k}^+, \quad B_k = \frac{1}{\sqrt{2}} \left( b_k - b_{-k}^+ \right) = -B_{-k}^+$$

$$C_\kappa = \frac{1}{\sqrt{2}} \left( a_\kappa + a_{-\kappa}^+ \right) = C_{-\kappa}^+, \quad D_\kappa = \frac{1}{\sqrt{2}} \left( a_\kappa - a_{-\kappa}^+ \right) = -D_{-\kappa}^+ \tag{G2}$$

The Hamiltonian Eq. (G1) can be rewritten as

$$H = \frac{1}{2}\sum_k \omega_k \left(A_k^+ A_k + B_k^+ B_k\right) + \frac{1}{2}\sum_\kappa \Omega(\kappa)\left(C_\kappa^+ C_\kappa + D_\kappa^+ D_\kappa\right) + H_{int}$$

$$H_{int} = \sum_k M_k B_k^+ D_\kappa$$

(G3)

Now defining phonon and other relevant Green's functions as

$$G_{kq}(t-t') = -i\theta(t-t')\langle[A_k(t), A_q^+(t')]\rangle, \quad B_{kq}(t-t') = -i\theta(t-t')\langle[B_k(t), A_q^+(t')]\rangle$$

$$C_{\kappa q}(t-t') = -i\theta(t-t')\langle[C_\kappa(t), A_q^+(t')]\rangle, \quad D_{\kappa q}(t-t') = -i\theta(t-t')\langle[D_\kappa(t), A_q^+(t')]\rangle$$

(G4)

Now take time derivative to Eq. (G4), noticing that $[H_0, A_k] = -\omega_k B_k$, $[H_{int}, A_k] = -M_k D_\kappa$, $[H_0, B_k] = -\omega_k A_k$, $[H_{int}, C_\kappa] = \sum_p M_p B_p^+ \delta_{\kappa,-p_z} = -\sum_p M_{-p} B_p \delta_{\kappa,p_z}$, $[H_0, C_\kappa] = -\Omega(\kappa) D_\kappa$, we have

$$i\partial_t G_{kq}(t-t') = \omega_k B_{kq}(t-t') + M_k D_{\kappa q}(t-t')$$

$$i\partial_t B_{kq}(t-t') = \delta(t-t')\delta_{kq} + \omega_k G_{kq}(t-t')$$

$$i\partial_t C_{\kappa q}(t-t') = \Omega_\kappa D_{\kappa q}(t-t') + \sum_p M_{-p} \delta_{\kappa,p_z} B_{pq}(t-t')$$

$$i\partial_t D_{\kappa q}(t-t') = \Omega_\kappa C_{\kappa q}(t-t')$$

(G5)

Performing Fourier transform, we obtain

$$\omega G_{kq}(\omega) = \omega_k B_{kq}(\omega) + M_k D_{\kappa q}(\omega)$$

$$\omega B_{kq}(\omega) = \frac{1}{2\pi}\delta_{kq} + \omega_k G_{kq}(\omega)$$

$$\omega C_{\kappa q}(\omega) = \Omega_\kappa D_{\kappa q}(\omega) + \sum_p M_{-p}\delta_{\kappa p_z} B_{pq}(\omega)$$

$$\omega D_{\kappa q}(\omega) = \Omega_\kappa C_{\kappa q}(\omega)$$

(G6)

Solving Eq. (G6), we obtain the self-consistent Dyson equation of phonon propagator

$$G_{kq}(\omega) = \frac{1}{2\pi}\frac{\omega_k}{\omega^2 - \omega_k^2}\delta_{kq} + \frac{M_k \Omega_\kappa}{(\omega^2 - \omega_k^2)(\omega^2 - \Omega_\kappa^2)}\left(\frac{1}{2\pi}M_{-q}\delta_{\kappa q_z} + \sum_p M_{-p}\delta_{\kappa p_z}\omega_p G_{pq}(\omega)\right)$$

(G7)

To first order, the $G_{pq}(\omega)$ on the right hand side are approximated as

$G_{pq}(\omega) = G_{pq}^0(\omega) = \frac{1}{2\pi}\frac{\omega_p}{\omega^2 - \omega_p^2}\delta_{pq}$. Hence Eq. (G7) is highly divergent when $\omega = \omega_k = \Omega_\kappa$, giving a rough resonance condition. In fact, Eq. (G7) can be solved exactly since the coefficient has no true κ-dependence as

$$G_{kq} = A_{kq} + B_k \sum_p C_p A_{pq} \bigg/ \left(1 - \sum_p C_p B_p\right)$$

(G8)

where $A_{kq} \equiv \frac{1}{2\pi} \frac{\omega_k}{\omega^2 - \omega_k^2} \delta_{kq} + \frac{1}{2\pi} M_{-q} \delta_{\kappa q_z} \frac{M_k \Omega_\kappa}{(\omega^2 - \omega_k^2)(\omega^2 - \Omega_\kappa^2)}$, $B_k \equiv \frac{M_k \Omega_\kappa}{(\omega^2 - \omega_k^2)(\omega^2 - \Omega_\kappa^2)}$,

$C_p \equiv M_{-p} \delta_{\kappa p_z} \omega_p$.

## H. Kubo formula for thermal conductivity in Matsubara frequency

The usual Kubo formula for bulk thermal conductivity $k$ can be written as [S6]

$$k = \frac{k_B \beta}{3V} \lim_{\delta \to 0} \int_0^{+\infty} e^{-\delta t} dt \int_0^\beta d\lambda \langle S(0) \cdot S(t+i\lambda) \rangle \tag{H1}$$

where $\beta \equiv 1/k_B T$, S is the energy flow operator which can be written as $S(t) = \sum_k v_k E_k n_k(t)$ with $v_k$ the group velocity vector, $n_k(t)$ is the number density operator [S7], $\langle \hat{O} \rangle = \frac{\text{Tr} e^{-\beta H} \hat{O}}{\text{Tr} e^{-\beta H}} = \frac{\int D(\bar{\varphi}, \varphi) \hat{O} e^{-S_{\text{eff}}}}{\int D(\bar{\varphi}, \varphi) e^{-S_{\text{eff}}}}$ is the thermodynamic average. Using Wick's theorem, and noticing the fact that neither non-conserving term $\langle \bar{\varphi}\bar{\varphi} \rangle$ nor equal-time (number density) $\langle \bar{\varphi}_k(0) \varphi_k(0) \rangle$ term contribute to transport property, we have

$$k = \frac{k_B \beta}{3V} \sum_{kp} v_k \cdot v_p E_k E_p \lim_{\delta \to 0} \int_0^{+\infty} e^{-\delta t} dt \int_0^\beta d\lambda \langle \bar{\varphi}_k(0) \varphi_p(t+i\lambda) \rangle \langle \varphi_k(0) \bar{\varphi}_p(t+i\lambda) \rangle \tag{H2}$$

The imaginary time Green's function can be written as

$$G_{nkp} = \frac{1}{2} \int_{-\beta}^{\beta} e^{-i\omega_n \tau} G_{kp}(\tau) d\tau$$

$$G_{kp}(\tau) = \frac{1}{\beta} \sum_n G_{nkp} e^{+i\omega_n \tau} = \langle T_\tau [\varphi_k(\tau) \bar{\varphi}_p(0)] \rangle \tag{H3}$$

Where $G_{nkp} \equiv \langle \varphi_{nk} \bar{\varphi}_{mp} \rangle = \frac{\delta_{mn} G_{0nk}}{\frac{1}{2} \delta_{pk} - J_{n\kappa} G_{0nk} \delta_{\kappa p_z} g_k^* g_p + \sqrt{\frac{1}{4} \delta_{pk} + (J_{n\kappa} G_{0nk} \delta_{\kappa p_z} g_k^* g_p)^2}}$ is the Matsubara Green's function obtained directly from Eq. (D11). The rest is to connect real time to Matsubara formulism.

It can be proven that the real-time commutator can be expressed in Lehmann representation as

$$\langle \bar{\varphi}_p(0) \varphi_k(t) \rangle = \frac{1}{Z} \sum_{mn} e^{-\beta E_m} \langle n|a_k|m \rangle \langle m|a_p^+|n \rangle e^{-i(E_m - E_n)t}$$

$$\langle \varphi_k(t) \bar{\varphi}_p(0) \rangle = \frac{1}{Z} \sum_{nm} e^{-\beta E_n} \langle n|a_k|m \rangle \langle m|a_p^+|n \rangle e^{-i(E_m - E_n)t} \tag{H4}$$

Now define spectral function as

$$A_{kp}(\omega) \equiv \frac{1}{Z}\sum_{mn} e^{-\beta E_n} \langle n|a_k|m\rangle\langle m|a_p^+|n\rangle \delta(\omega + E_m - E_n) \tag{H5}$$

We have

$$\langle \varphi_k(t)\bar{\varphi}_p(0)\rangle = \int_{-\infty}^{+\infty} A_{kp}(\omega) e^{+i\omega t} d\omega$$

$$\langle \bar{\varphi}_p(0)\varphi_k(t)\rangle = \int_{-\infty}^{+\infty} A_{kp}(\omega) e^{\beta\omega} e^{+i\omega t} d\omega \tag{H6}$$

Then the thermal conductivity can further be reduced using Eq. (H6) as

$$k = \frac{\pi k_B \beta^2}{3V} \sum_{kp} \mathbf{v}_k \cdot \mathbf{v}_p E_k E_p \int_{-\infty}^{+\infty} d\omega A_{kp}(\omega) A_{pk}(\omega) \tag{H7}$$

Where we have used the fact that Eq. (H2) is invariant under the transform $k \leftrightarrow p$, $\omega \leftrightarrow \omega'$.

Since the Matsubara Green's function can also be expressed in Lehmann representation

$$G_{skp} = \frac{1}{Z}\sum_{nm} \frac{\langle n|a_k|m\rangle\langle m|a_p^+|n\rangle}{E_n - E_m - i\omega_s}\left(e^{-\beta E_m} - e^{-\beta E_n}\right) \tag{H8}$$

From which we could obtain restarted and advanced Green's function through analytical continuation as $G_{kp}^R(\omega) = G_{skp}(i\omega_s \to \omega + i\delta)$ and $G_{kp}^A(\omega) = G_{skp}(i\omega_s \to \omega - i\delta)$, then combining Eq. (H5), it can be proven that

$$G_{kp}^R(\omega) - G_{kp}^A(\omega) = 2\pi i \left(e^{\beta\omega} - 1\right) A_{kp}(\omega) \tag{H9}$$

When substituting Eq. (H9) to Eq. (H7), we have

$$k(T) = -\frac{k_B \beta^2}{12\pi V}\sum_{kp} \mathbf{v}_k \cdot \mathbf{v}_p E_k E_p \int_{-\infty}^{+\infty} d\omega \frac{\left[G_{kp}^R(\omega) - G_{kp}^A(\omega)\right]\left[G_{pk}^R(\omega) - G_{pk}^A(\omega)\right]}{\left[\exp(\beta\omega) - 1\right]^2} \tag{H10}$$